\documentclass[aps,twocolumn,prl,amsmath,amssymb,floatfix,showpacs]{revtex4}

\newcommand{\op}[1]{\hat{#1}}
\newcommand{\opdag}[1]{\hat{#1}^{\dag}}

\newcommand{\bra}[1]{\mathinner{\langle{#1}|}}
\newcommand{\ket}[1]{\mathinner{|{#1}\rangle}}

\newcommand{\Tr}{\text{Tr}}

\usepackage{graphicx}
\usepackage{natbib}
\usepackage{multirow}

\begin{document}

\title{Ground-State Entanglement in a Coupled-Cavity Model}

\author{E. K. Irish}
\email{e.k.irish@leeds.ac.uk}
\affiliation{School of Physics and Astronomy, University of Leeds, Leeds LS2 9JT, United Kingdom}
\affiliation{School of Mathematics and Physics, Queen's University Belfast, Belfast BT7 1NN, United Kingdom}

\date{\today}

\begin{abstract}

Bipartite entanglement entropies are calculated for the ground state of the two-excitation subspace in a two-site coupled cavity model. Each region in the phase diagram (atomic insulator, polaritonic insulator, photonic superfluid, and polaritonic superfluid) is found to be characterized by unique entanglement properties. In particular, the polaritonic superfluid region exhibits multipartite entanglement among the two atoms and two cavity fields. This system provides a toy model in which a number of intriguing aspects of entanglement can be studied, such as the relationship of entanglement to phase transitions, entanglement of particles with different dimensionality, and the connection between experimentally accessible local observables and entanglement entropies. 

\end{abstract}


\maketitle

Quantum phase transitions have been studied for a number of years in condensed matter physics~\cite{Sachdev:QPT}. Unlike classical phase transitions, quantum phase transitions occur at zero temperature and are driven by quantum rather than thermal fluctuations. As a parameter is varied through the quantum critical point, the ground state of the system undergoes a transition characterized by a change in the correlations among particles. For example, the Bose-Hubbard model exhibits a transition from a product of localized states of definite particle number (Mott insulator) to a delocalized state with large fluctuations in the particle number per site (superfluid)~\cite{Fisher:1989}.

Examining the nature of these correlations led to the idea that entanglement could be a useful tool for understanding the complex collective phenomena involved in quantum phase transitions~\cite{Osterloh:2002,Osborne:2002,Vidal:2003,Amico:2008,Farias:2008,Goold:2009}. This insight has been employed quite successfully in spin-$\tfrac{1}{2}$ systems, although other types of systems are more difficult to treat within the current theory of entanglement~\cite{Johannesson:2007,Amico:2008}.

Recently it has been shown that an array of optical cavities, where each cavity contains one or more atoms, can undergo a quantum phase transition~\cite{Hartmann:2006,Greentree:2006,Angelakis:2007b,Rossini:2007,Hartmann:2007c,Huo:2008,Irish:2008,Hartmann:2008,Zhao:2008,Aichhorn:2008}. Models of this type can be effectively described by the Bose-Hubbard model in some limits~\cite{Hartmann:2006,Greentree:2006,Angelakis:2007b,Rossini:2007,Huo:2008}. However, the particles involved are not bosons but polaritons, quasiparticles composed of an atom plus photons. This difference adds a new dimension to the problem~\cite{Irish:2008,Zhao:2008,Ogden:2008,Aichhorn:2008}.

Already there has been some work on the emergence of entanglement in coupled-cavity systems. Most of it has focused on the dynamic production of various entangled states for applications in quantum information processing. Proposals for creating entanglement between two polaritons~\cite{Angelakis:2007c}, two photons~\cite{Angelakis:2007a,Cho:2008}, two atoms~\cite{Cho:2008}, or a photon and an atom~\cite{Cho:2008} have appeared. There have also been proposals for creating highly entangled cluster states of polaritons~\cite{Hartmann:2007a,Angelakis:2008}. Huo et al.~\cite{Huo:2008} compared the atom-atom entanglement and the visibility of photon interference fringes with the fluctuation of the total excitation number as means of characterizing the quantum phase transition. They found that the phase transition affects the entanglement between two atoms.

In this paper I examine the ground-state entanglement in a simple two-site coupled-cavity model and how it relates to the various states in the phase diagram. Each site consists of a single-mode cavity containing a two-level atom; the two sites are identical. The cavities are coupled so that allows photons can hop between them. Although it is not strictly correct to refer to ``phases'' in such a small system, it has been shown that many results from finite models can be extrapolated to the thermodynamic limit~\cite{Amico:2008}. Crucially, in a system of this size it is possible to calculate the entanglement entropy of all the possible bipartitions of the ground state. Such detailed examination rapidly becomes more difficult as the size of the system increases. Furthermore, models with only a few sites could easily be implemented in superconducting qubit or quantum dot experiments~\cite{Wallraff:2004,Majer:2007,Benyoucef:2008}.

The Hamiltonian for the model is given by ($\hbar = 1$)
\begin{equation}\label{hamiltonian}
\begin{split}
H &=  \sum_{j=1,2}[\omega_c \opdag{a}_j \op{a}_j + \omega_a\ket{e_j} \bra{e_j} \\
& \quad \quad \quad + g(\opdag{a}_j \ket{g_j}\bra{e_j} + \op{a}_j \ket{e_j}\bra{g_j})] + A(\opdag{a}_1 \op{a}_2 + \opdag{a}_2 \op{a}_1) ,
\end{split}
\end{equation}
where $\omega_c$ and $\omega_a$ are the cavity and atom frequencies, $g$ is the atom-cavity coupling strength, and $A$ is the hopping strength. The operator $\op{a}_j$ ($\opdag{a}_j$) is the lowering (raising) operator for the field in cavity $j$. The states $\ket{g_j}$ and $\ket{e_j}$ represent the ground and excited states of the atom in cavity $j$. Hence the operator $\ket{g_j}\bra{e_j}$ ($\ket{e_j}\bra{g_j}$) is the atomic lowering (raising) operator.

A useful property of Eq.~\eqref{hamiltonian} is that the total number of excitations $\op{N} = \opdag{a}_1 \op{a}_1 + \opdag{a}_2 \op{a}_2 + \ket{e_1} \bra{e_1} + \ket{e_2} \bra{e_2}$ is conserved. For simplicity, the analysis is restricted to the case of two excitations. 

In the limit of zero hopping ($A=0$), the system reduces to a pair of noninteracting atom-cavity systems, each described by the Jaynes-Cummings model~\cite{Jaynes:1963,Shore:1993}. The usual rotating-wave approximation has been applied, so that the interaction between the atom and the field contains only ``energy-conserving'' terms. This model is exactly solvable; the eigenstates of the individual cavities are given by
\begin{gather}
\ket{0_i} = \ket{g_i}\ket{0_i}, \label{polariton0}\\
\ket{n_i^-} = \sin \frac{\theta_n}{2} \ket{e_i}\ket{(n-1)_i} - \cos \frac{\theta_n}{2} \ket{g_i}\ket{n_i}, \label{polaritonminus} \\
\ket{n_i^+} = \cos \frac{\theta_n}{2} \ket{e_i}\ket{(n-1)_i} + \sin \frac{\theta_n}{2} \ket{g_i}\ket{n_i}, \label{polaritonplus}
\end{gather}
where $i=1,2$ denotes the site, $\ket{n}$ ($n=1,2,3,\dots$) is a photon number state, and $\tan \theta_n = 2 g \sqrt{n}/\Delta$, where $\Delta = \omega_a - \omega_c$ is the detuning. 

The excited states are known as dressed states or polaritons, and their nature depends on the atom-cavity detuning $\Delta$. Consider, for example, the state $\ket{1_i^-}$. At exact resonance ($\Delta = 0$), it is a maximally entangled state of an atom and a single photon. As $\Delta \to +\infty$, $\ket{1_i^-} \to \ket{g_i}\ket{1_i}$: that is, the excitation becomes purely photonic. Likewise, as $\Delta \to -\infty$, the excitation becomes purely atomic. 

Previous work has shown that the coupled-cavity model can undergo a phase transition similar to that of the Bose-Hubbard model~\cite{Hartmann:2006,Greentree:2006,Angelakis:2007b}. Near resonance the atom-cavity interaction shifts the frequency of the cavity, causing the ``photon blockade'' effect~\cite{Imamoglu:1997,Birnbaum:2005} that prevents an additional photon from entering the cavity. The result is an effective on-site repulsion, as in the Bose-Hubbard model. For small values of $A$, this effective potential enforces the insulatorlike character of the ground state, characterized by a well-defined number of excitations at each site. As $A$ increases, the hopping is able to overcome the repulsion. The ground state becomes delocalized and the local excitation number is no longer fixed, similar to a superfluid state. Since only photons are able to hop between cavities, the limiting state as $A$ becomes large may be characterized as a photonic superfluid.

\begin{figure}
\includegraphics[scale=1]{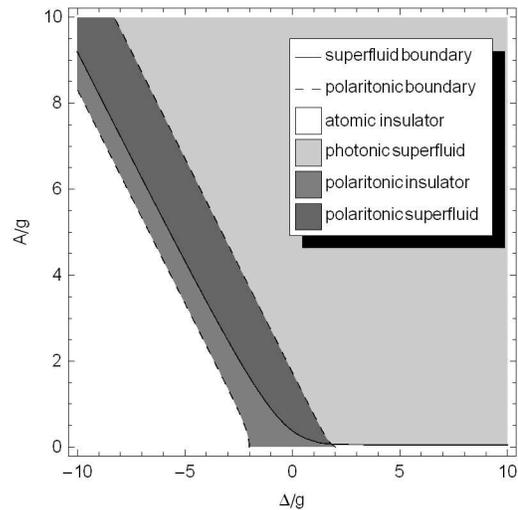}
\caption{\label{fig:phasediagram} Phase diagram of the two-cavity system. The phase boundaries are defined as one-half the maximum value of the appropriate order parameter. For the superfluid state the variance of the total excitation number at site 1 is used as the order parameter, while for the polaritonic state the variance of the atomic excitation number at site 1 is used~\cite{Irish:2008}.}
\end{figure}

Figure~\ref{fig:phasediagram} shows a phase diagram of the coupled cavity system. Four different regions have been identified, distinguished by the phase (insulator or superfluid) and the nature of the particles involved~\cite{Irish:2008}. The insulator state may be polaritonic (for small $\Delta$) or atomic (for large negative $\Delta$). As $A$ becomes large, the ground state becomes a photonic superfluid. Perhaps the most interesting region, however, appears in the vicinity of the transition around $A = -\Delta$. The state in this region has both superfluid and polaritonic characteristics.

The phase diagram shown in Fig.~\ref{fig:phasediagram} was obtained, as in Ref.~\cite{Irish:2008}, by considering the variances of the total and atomic excitation numbers at a single site. For the Bose-Hubbard model, it has been shown that the entanglement properties of the ground state mirror the behavior of the number variance in signaling the phase transition~\cite{Farias:2008}. Similarly, it seems reasonable to expect that the phases in the coupled-cavity model may be distinguished by their entanglement properties.

There are two sources of entanglement in the coupled cavity system. The first is the polaritonic entanglement due to the atom-cavity interaction. From Eqs.~\eqref{polaritonminus}-\eqref{polaritonplus} it may be seen that the degree of entanglement of the polaritonic states is controlled by the interaction strength $g$ and the detuning $\Delta$. The polaritonic entanglement is local in character in that it occurs between the atom and the cavity field at a single site only. Thus it is not included in entanglement measures that treat the interacting atom and field at one site as a single ``particle''. The second source of entanglement is the phase transition, controlled by the hopping parameter $A$. Phase-transition-induced entanglement may appear at any value of $\Delta$, although the value of $A$ at which the transition occurs depends on $\Delta$. This type of entanglement is generally nonlocal in character.

In this paper I focus on the bipartite entanglement in the ground state of the system, which is of course a pure state. Accordingly, the von Neumann entropy is a good entanglement measure. If a pure state of a composite system $AB$ is given by the density matrix $\rho_{AB}$, the entropy $S(\rho_A) \equiv -\Tr(\rho_A \log_2 \rho_A)$ of the reduced density matrix $\rho_A = \Tr_B (\rho_{AB})$ measures the entanglement between part $A$ and part $B$ of the system.

\begin{figure}
\includegraphics[scale=1]{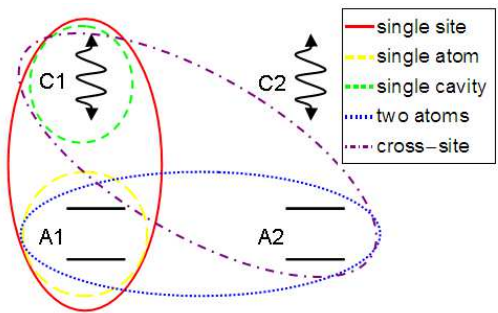}
\caption{\label{fig:bipartitions} Illustration of the five different bipartitions in the two-site system. }
\end{figure}

Assuming that the two sites are identical, there are five possible bipartitions, as illustrated in Fig.~\ref{fig:bipartitions}. Each bipartition corresponds to an entropy measure. The single site entropy $S(\rho_{A1C1})$ is found by tracing over the atom and cavity of site 2 and applying the entropy formula to the resulting reduced density matrix. This entropy is essentially insensitive to the polaritonic entanglement. It therefore provides a measure of the nonlocal entanglement. $S(\rho_{A1})$ and $S(\rho_{C1})$ give the entanglement of a single atom and a single cavity, respectively. We must also look at the entropy of both atoms $S(\rho_{A1A2})$ (equal to the entropy of both cavities) and the cross-site entropy $S(\rho_{A1C2})$. 

When comparing the various entropies, it is important to note that the von Neumann entropy has a maximum value of $\log_2 d$ in a $d$-dimensional space. For a single two-level atom, $d=2$ and $S_{max}(\rho_{A1})=1$. Although the dimension of a single cavity field is infinite, the restriction to two excitations means that each field forms an effective three-level system, so $S_{max}(\rho_{C1})=\log_2 3 \approx 1.585$. The atomic subspace has dimension $4$, giving $S_{max}(\rho_{A1A2})=2$. The single-site and cross-site bipartitions each contain one atom and one cavity field for a nominal dimension of $6$. However, the restriction to two excitations limits the number of atom-cavity states to $5$, which corresponds to $S_{max}(\rho_{A1C1})=S_{max}(\rho_{A1C2})=\log_2 5 \approx 2.322$.

\begin{figure}
\includegraphics[scale=1]{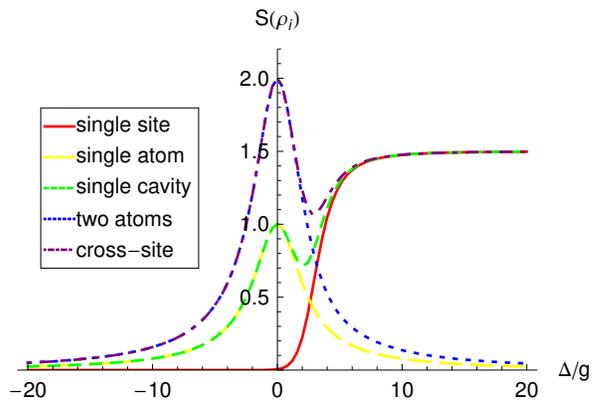}
\caption{\label{fig:entropies1} Bipartite entropies as a function of $\Delta$ in the small-hopping limit, $A=0.01g$.}
\end{figure}

Figure~\ref{fig:entropies1} shows the five bipartite entropies in the limit of small hopping. At large negative detuning all of the entropies are nearly zero, as expected from the characterization of this region as an atomic insulator. Near $\Delta = 0$, four of the five entropies reach a local maximum, reflecting the polaritonic entanglement created by the resonance between the atom and the cavity field at each site. However, the single site entropy remains close to zero and thus there is no intersite entanglement. This corresponds to a polaritonic insulator state, which takes the form of a tensor product across sites where each site contains an entangled state. Finally, as $\Delta$ increases, the single atom and atomic subspace entropies drop to zero while the other three entropies asymptotically approach $1.5$. In this photonic superfluid region, the atoms are both in the ground state. The photons in the two sites are entangled, as evidenced by the nonzero single site, single cavity, and cross site entropies (which become equivalent as $\Delta \to \infty$).

\begin{figure}
\includegraphics[scale=1]{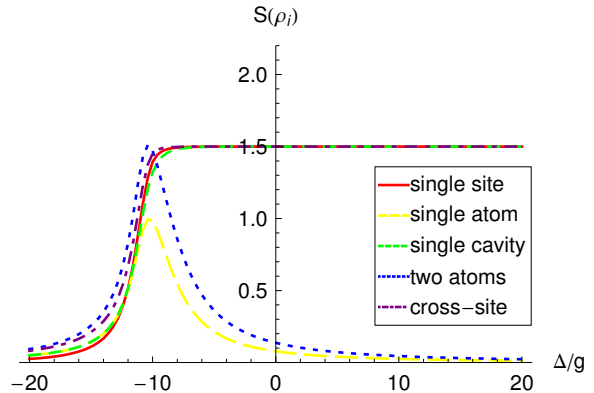}
\caption{\label{fig:entropies3}Bipartite entropies as a function of $\Delta$ in the large-hopping limit, $A=10g$. }
\end{figure}

The term ``polaritonic superfluid'' suggests a delocalized state composed of polaritons, which are themselves entangled states of an atom and a cavity field. This suggests that the polaritonic superfluid state might exhibit true multipartite entanglement. Although a multipartite entanglement measure for four particles of mixed dimensionality does not yet exist to our knowledge, an indicator of multipartite entanglement is that all possible bipartitions of the system are entangled. A plot of the entropies in the large-hopping limit is shown in Fig.~\ref{fig:entropies3}. As in the small-hopping limit, the state at large negative detuning is an atomic insulator, while at large positive detuning the state is a photonic superfluid. However, all five bipartite entropies reach a maximum near the transition point $-\Delta \approx A$, indicating that there is indeed multipartite entanglement in the polaritonic superfluid state. 

Note that although only two extreme cases typifying the small- and large-hopping limits are shown, the entanglement peaks follow the transition throughout the whole parameter space. At intermediate values of $A$, the entropy curves transform between the two limiting cases in a predictable fashion.

It should also be pointed out that the multipartite entanglement in this system only occurs because the atom-cavity structure allows multiple subsystems at each lattice site. Other authors have considered the relation between phase transitions and multipartite entanglement among more than two lattice points~\cite{Anfossi:2005,deOliveira:2006,Costantini:2007,Farias:2008}, but that is a somewhat different problem.

Entanglement properties, therefore, provide a useful means of differentiating the various regions in the phase diagram. The atomic insulator state is completely separable. In the polaritonic insulator state, polaritonic entanglement exists between the atom and field at each site, but there is no intersite entanglement. The photonic superfluid state exhibits bipartite entanglement between the sites. Genuine multipartite entanglement appears in the polaritonic superfluid state. This characterization suggests a connection between the entropies and the number variances that have served as order parameters in previous work, which is currently being investigated. However, it seems that no single bipartite entropy provides a complete view of the system; they need to be taken together for a full description.

Furthermore, the description in terms of entanglement reveals a qualitative difference between the transitions at large and small hopping values. In the small hopping limit, the change in particle type from atomic to polaritonic to photonic is due to the resonance between the atoms and photons. The phase transition occurs at a larger value of $\Delta$ than the resonance, so the region of polaritonic behavior does not overlap much with the superfluid region. As a consequence, the entanglement remains primarily bipartite. In contrast, for large hopping values the phase transition occurs deep in the negative detuning regime. The change in particle type is driven by the transition itself, resulting in multipartite entanglement in the transition region. Whether this difference, and indeed the polaritonic superfluid state itself~\cite{Zhao:2008}, persists in the thermodynamic limit remains an open question.

The model studied here, although itself quite simple, opens up some significant avenues of inquiry. It provides a toy model for the problem of entanglement of particles with different dimensionality. The interplay between on-site and intersite entanglement is another interesting problem: how is entanglement created, transformed, and destroyed in the system? Of course, there are also questions about practical applications, particularly whether this model can be used to produce useful entanglement for tasks such as computation. Furthermore, the connection between the variances and entropies could potentially offer an experimentally accessible way to measure entanglement~\cite{Klich:2006}.

\acknowledgments

I would like to thank M.S. Kim, C.D. Ogden, M. Paternostro, and V. Vedral for helpful discussions and suggestions. Support from DEL, UK EPSRC, and QIPIRC is gratefully acknowledged.

\bibliography{../../../masterbib}

\end{document}